\documentclass[twocolumn,showpacs,preprintnumbers,
amsmath,amssymb,aps,prd,nofootinbib,superscriptaddress,
eqsecnum]{revtex4}
\usepackage{graphicx}
\usepackage{dcolumn}
\makeatletter
\renewcommand{\p@subsection}{}
\makeatother

\newcommand{\Rmnum}[1]{\expandafter\@slowromancap\romannumeral #1@}

\newcommand{\be}{\begin{eqnarray}}
\newcommand{\ee}{\end{eqnarray}}

\def\lsim{\mathrel{\rlap{\lower3pt\hbox{\hskip1pt$\sim$}}
     \raise1pt\hbox{$<$}}} 
\def\gsim{\mathrel{\rlap{\lower3pt\hbox{\hskip1pt$\sim$}}
     \raise1pt\hbox{$>$}}} 

\def\la{\langle}
\def\ra{\rangle}
\def\bi{\bibitem}

\begin{document}

\title{Proton Mass, Topology Change and Tensor Forces\\ in Compressed Baryonic Matter}


\author{Mannque Rho}

\affiliation{%
Institut de Physique Th\'eorique,
CEA Saclay, 91191 Gif-sur-Yvette c\'edex, France \& \\
Department of Physics, Hanyang University, Seoul 133-791, Korea
}

\date{\today}

\begin{abstract}
This is a summary of the talks I gave at Korean Physical Society meeting (April 26, 2012, Daejeon, Korea) and the 4th  Asian Triangle Heavy Ion Conference (ATHIC) (November 14, 2012, Pusan, Korea). They are based on the series of work done at Hanyang University in the World Class University III Program under the theme of ``From Dense Matter to Compact Stars." The program was conceived and executed to understand highly compressed baryonic matter in anticipation of the forthcoming RIB machine ``RAON" which is in construction in the Institute for Basic Science (IBS) in Korea. The problems treated ranged from the origin of the proton mass,  topological structure of barynic matter,  chiral symmetry and conformal symmetry to the EoS of nuclear matter and dense neutron-rich matter and to the maximum mass of neutron stars. The results obtained are new and intriguing and could have an impact on the novel structure of dense matter to be probed in the accelerators ``RAON," FAIR etc. and in compact stars.

\end{abstract}

\pacs{}

\maketitle


\section{Introduction}
The landscape of hadronic phases has been fairly extensively explored at high temperatures thanks to lattice QCD on theory side and to RHIC and now LHC on experimental side, but it is a totally barren uncharted field in the direction of high density at low temperature. Lattice QCD cannot access the density regime relevant to the interior of compact stars, at present the only source available for high density, and there are no theoretical tools that have been confirmed reliable, given the lack of experiments available at high density. The phase structure of compressed baryonic matter beyond nuclear matter density is more or less unknown and poses a challenge in hadron/nuclear physics.

In this note I would like to discuss a line of work done recently to go from nuclear matter density to densities relevant to the interior of compact stars at zero temperature. The merit of the work is that it is a unified approach to baryons and mesons in and out of medium anchored on one effective Lagrangian with the symmetries (assumed to be) embodied in QCD, valid up to the density at which deconfinement sets in. The principal theme will be that the origin of the proton mass plays a key role in the EoS for compact stars and can be explored in the forthcoming terrestrial accelerators such as RIB machines (e.g., ``KoRIA" or RAON in Korea, FRIB of MSU/Michigan ...), FAIR of Darmstadt/Germany, NICA at Dubna/ Russia etc. and the space observatories in operation and in project. I suggest this as a direction for the coming era in hadron/nuclear physics in Korea.

The problem is that while quark masses could be explained by the discovery of the Higgs or Higgs-like boson, the bulk, say, 98\%, of the mass of proton whose constituents are nearly zero mass quarks and massless gluons,  remains unexplained. This is because unlike molecules, atoms and nuclei, the mass of the proton is a ``mass without mass."\cite{wilczek}.

In the standard lore, the proton mass (and the mass of  ``light-quark mesons,"  say, the $\rho$ meson to be specific), if one assumes that the up and down quark masses are zero (called the chiral limit), is said to be entirely ``generated dynamically."  Phrased in terms of symmetries, the mass is attributed to the spontaneous breaking of chiral symmetry (SBCS for short). The SBCS is characterized by that the quark condensate $\la\bar{q}q\ra$, the vacuum expectation value (VEV) of the bilinear quark fields, is nonvanishing, $\la\bar{q}q\ra_0\neq 0$. If this were the entire story, then one would have,  by `dialling' the quark condensate, that
\be
m_N(\la\bar{q}q\ra)\rightarrow 0 \ \ {\rm as}\ \ \la\bar{q}q\ra\rightarrow 0.
\ee
But QCD does not say that this is the entire story. In fact, it is possible to have a mass term in the proton that does not vanish when the quark condensate goes to zero without violating chiral symmetry in the chiral limit. It is easy to undersand this if one recalls the $SU(2)\times SU(2)$ Gell-Mann-L\'evy linear sigma model with the doublet nucleons, the triplet pions and the scalar $\sigma$ meson~\cite{gell-mann-levy}. In this model, the nucleon and the scalar $\sigma$ acquire masses entirely by the VEV of the sigma field $\la\sigma\ra_0\neq 0$ while the pion remains massless by Nambu-Goldstone theorem. On the other hand, as noted by DeTar and Kunihiro~\cite{detar}, one can have the nucleon mass in the form
\be
m_N=m_0 +\bar{m} (\la\bar{q}q\ra)\label{Nmass}
\ee
such that $\bar{m}\rightarrow 0$ as the condensate is dialled to zero {\em provided} one introduces parity doublet to the nucleon. Then the nucleon mass does not vanish if $m_0$ does not. Here $m_0$ is a chirally-invariant mass term. This model is referred to as baryon parity-doublet model.

Now what about the mesons? In the framework I am adopting, the situation is different for mesons. When the chiral symmetry is restored, the $\sigma$ mass has to join the pion mass, hence must go to zero in the chiral limit. When vector mesons are suitably introduced according to hidden local symmetry which is gauge equivalent to the (nonlinear) sigma model~\cite{HY:PR}, the mass of the vector meson $\rho$ also has to go to zero due to what is known as ``vector manifestation" (VM for short)  of hidden local symmetry. It may be that the masses of other light-quark mesons belonging to certain flavor symmetries all go to zero in the sense of ``mended symmetries" \`a la Weinberg~\cite{weinberg}. There is thus an apparent difference in this picture between baryon and meson masses in the way chiral restoration is reached. I will discuss later that this difference can be avoided if one resorts to the constituent quark picture. I should note in this connection that by artificially unbreaking chiral symmetry in a dynamical lattice simulation~\cite{Lang-unbreaking}, Glozman et al. find that both mesons and baryons remain massive after chiral symmetry is presumably restored~\cite{glozman}. In fact, in baryons, $m_0$ is found to be large.

The underlying theme in my discussion will be that a substantial $m_0$ is indicated in the equation of state (EoS) of nuclear matter and dense neutron-rich (compact-star) matter. It figures in nuclear dynamics in a highly intricate way. It is important to note that such a large $m_0$ raises a fundamental question in physics: Where does the proton mass associated with $m_0$ come from if it is unconnected to the spontaneous breaking of chiral symmetry?
\section{Unbreaking symmetries}
If hadron masses arise by the spontaneous breaking of chiral symmetry by the vacuum characterized by the condensate $\la\bar{q}q\ra\neq 0$, then one should be able to tweak the vacuum such that the broken symmetry is restored and the fate of hadron masses is revealed. It is generally believed that when a hadronic system is heated to high enough temperature or compressed to high enough density, the quark  condensate will go to zero or to near zero. This is confirmed by lattice QCD in the chiral limit for high temperature. The situation with the density, however,  is totally unclear because there is no QCD calculation for dense matter: Lattice method cannot access high-density matter. So it is not known rigorously whether the quark condensate does indeed vanish at some high density or put another way, whether  chiral symmetry is actually restored at such high density presumably present in the interior of compact stars. One would like to find this out by experiments.

Numerous efforts have been made -- and will continue to be made -- to verify whether the presumed link between the mass and the quark condensate can be established. One obvious experimental way is to try to ``unbreak chiral symmetry" by heating hadronic matter to high temperature or compressing it to high density by heavy ion collisions with the objective to make the quark condensate go to zero. So far there have been many (elaborate) experiments to do just this, but I have to say that up to date no convincing evidence, either positive or negative, has come up. No experiments so far succeeded to single out unambiguously the order parameter that indicates what is happening. Let me briefly explain this without going into details of the experiments performed and the numerous theoretical interpretations put forward `explaining' the observations.

The currently popular idea for unravelling vacuum structure of chiral symmetry is to `measure' the mass of a hadron in a medium, at high temperature and/or at high  density. But the question is: What is the meaning of ``in-medium" mass? Since the detector is outside of the medium where the temperature and density have vacuum values while the quantity of a hadron one wants to measure is the one inside a medium, there is no unambiguous meaning of mass inside hot or dense medium when measured on the detector located outside. In the past, what one did was to detect weakly interacting particles that carry with as little disturbance as possible the snap-shot picture of the interactions that take place in the medium. Thus one measured the invariant mass of the lepton pairs $l^+ l^-$ where $l=e, \mu$ arising from the decay of in-medium vector mesons, typically the $\rho$ meson.  The hope was to see the invariant masses corresponding to the $\rho$ meson mass sliding in medium with densitiy or temperature reflecting the in-medium ``vacuum" property of the quark condensate. These were studied in heavy-ion collisions probing high temperature and in the electroproduction of $\rho$ mesions in nuclei probing density effects. It turned out that the results were inconclusive. In \cite{blind}, I give my arguments chiefly drawn from the references given in \cite{blind} that those experiments did not succeed to single out, and take snapshot of, the order parameter $\la\bar{q}q\ra$ as intended.

That the vector meson mass sliding with density or with temperature measured in dilepton productions could not give a good snapshot of the vacuum structure reflecting chiral symmetry was already pointed out right after the scaling relation -- called ``BR scaling"~\cite{BR91} -- was proposed. See \cite{KR} for a footnote remark on this point.

There are several reasons why such measurements could have failed to exhibit the phenomenon looked for. The most trivial -- and unquestionably disappointing -- possibility I did not touch on in \cite{blind} could be that as one approaches the chiral restoration point at high temperature and/or high density, hadrons could simply break into pieces and cannot be described by local fields. The class of pictures where percolation takes place before the transition could perhaps be put in this category. Although this possibility cannot be dismissed -- and I have nothing further to say thereon, let me focus on the alternative possibility that hadronic degrees of freedom can be described in terms of local hadronic fields up to the point where the phase transition takes place.

If one assumes that local field theory makes sense up to the phase change,  then one can convince oneself, using hidden local symmetry, that certain properties of EM interactions that are known to hold in the (matter-free) vacuum do not apply to hot and/or dense matter. Of particular importance is that the ``vector dominance" -- which holds well in matter-free space and is simply assumed by most workers in the field to hold in medium -- does not hold as the temperature or the density approaches the critical.  The consequence discussed  in \cite{blind} is that the dileptons become  ``nearly blind" to the part of the mass, i.e., $\bar{m} (\la\bar{q}q\ra)$,  that encodes information on the quark condensate as it approaches zero.

As far as I know, the connection of the in-medium mass to the quark condensate $\la\bar{q}q\ra$ is precisely given only in hidden local symmetry theory and even there (in the chiral limit) only in the vicinity  of the VM fixed point with vanishing condensate. It has been established~\cite{HY:PR} that as $\la\bar{q}q\ra\rightarrow 0$, the pole mass -- not just the parametric mass -- of the vector meson $\rho$ goes as
\be
\frac{m^*_\rho}{m_\rho}\rightarrow \frac{g^*}{g}\propto \frac{\la\bar{q}q\ra^*}{\la\bar{q}q\ra}\label{VM}
\ee
where $g$ is the hidden gauge coupling. This comes about because of the flow to the VM fixed point in HLS theory (VM/HLS for short) that is forced upon by matching with QCD. I expect this to more or less hold near the chiral transition point even if the chiral limit is not assumed. In the close vicinity of the VM fixed point, the gauge coupling approaches zero so the width will also get suppressed. Thus the $\rho$ meson should become a sharper resonance near the critical point. The irony here is, however, that the photon becomes `nearly blind' to such dropping-mass vector mesons by the same VM mechanism~\cite{blind}.

Now most of the experimental measurements performed so far involve temperatures or densities remote from the critical point. In this case, the meaning of in-medium mass becomes even more blurred. In the hidden gauge theory framework that I am adopting, there is no reason for the direct link (\ref{VM}) between the mass and the condensate to hold far away from the VM fixed point. For instance in dense matter near nuclear matter density, the ``effective mass" of  the $\rho$ meson can be related to the Landau-Migdal Fermi-liquid parameter $F_1$, which is a fixed-point parameter in the effective field theory of nuclear matter~\cite{friman-rho}. An apt and non-trivial example of such a relation is found in the anomalous orbital gyromagnetic ratio $\delta g_l$ of heavy nuclei~\cite{friman-rho}. This relation means that the so-called density-dependent ``$\rho$ mass" contains, among others, quasiparticle interactions near the Fermi sea, a quantity that cannot be simply and directly linked to the order parameter of chiral symmetry.\footnote{It may very well be that the Landau parameters -- or in that matter, other mundane-looking nuclear effects such as collisional broadening discussed in the literature -- have something {\it ultimately} to do with chiral symmetry of QCD but arguing that they are consistent with or reflecting chiral symmetry would be senseless. It's as devoid of meaning as saying ``chiral dynamics explain nuclei."}  Stated more to the point, there is no way that one can single out the role of chiral symmetry in nuclear dynamics by `measuring' the property of a (light-quark) hadron in a process that takes place in a density regime near that of nuclear matter. The  presence of large widths etc. as observed in the experiments and much discussed in the literature should come as no surprise; it merely reflects that strong interactions are {\em undeed} taking place with a large number of  channels open in heat bath and/or compressed matter into which excitations of the $\rho$ quantum numbers can decay.  `Seeing' the $\rho$ meson signalling the effect of (\ref{VM}), even if present in such an environment, would be like seeing a needle in a haystack.

I will mention below that something analogous happens with the nuclear symmetry energy.
\section{Topology change}
I will now focus on density effects and consider dense baryonic matter. It will be described in terms of skyrmions that arise as topological solitons from an effective Lagrangian that has the symmetries assumed to be present in QCD.  There will be no need to put in baryons by hand.  As stated in Introduction, I will stick to a single effective Lagrangian, a generalized nonlinear sigma model that contains vector mesons \`a la hidden local symmetry (HLS) and a scalar meson, the dilaton $\chi$ associated with spontaneously broken scale symmetry (SBSS).\footnote{What is meant by dilaton is explained in Section \ref{dilaton}.} Let me call this dHLS Lagrangian, ``d" standing for the dilaton.  I will come back in the next section to the role of a scalar meson of vacuum mass $\sim 600-700$ MeV in nuclear interactions. It will play a crucial role on hadron masses sliding with density in effective Lagrangians.

The power of the skyrmion approach is that one can describe mesons, elementary baryons and multibaryon  matter, all with one effective Lagrangian. This allows one to do as consistent a treatment as feasible, avoiding arbitrary mixing of various different models in going from one regime of density to another regime as has been done in the past. I must admit that there is of course a price to pay for such a ``unified approach": Given the constraints and nonlinearity inherent in the Lagrangian picked,  it is technically difficult to do fully reliable quantum calculations. I will try -- and to some extent, succeed -- to finesse this difficulty by resorting to what Nature indicates, in particular in fixing parameters of the model. I will be generally thinking in terms of the dHLS Lagrangian $ {\cal L} (\pi,\rho,\omega,\chi)$\footnote{A work is in progress with a Lagrangian that contains an infinite tower of vector mesons coming from gravity-gauge dual QCD models~\cite{maetal}.}. However whenever qualitatively reliable,  I will discuss with the simpler Skyrme Lagrangian (containing the quadratic current algebra term and the quartic Skyrme term)~\cite{skyrme} implemented with the dilaton $\chi$ (call it dSkyrme). The dSkyrme can be considered as a dHLS from which the vector mesons are integrated out. It should be qualitatively reliable at densities far away from the VM fixed point where the hidden gauge  symmetry is indispensable and the vector mesons cannot be integrated out..

One natural way to describe many-baryon systems in the present framework is to put multi skyrmions on FCC crystal and reduce the crystal size to simulate dense matter~\cite{multifacet}. In the large $N_c$ consideration on which the skyrmion Lagrangian relies, it is justified to consider dense matter in a crystal form~\cite{klebanov}. But nuclear matter is most likely not in a crystal form, the deviation from crystal being effects higher order in $1/N_c$.  I will, however, argue that what we can reliably deduce from the crystal calculation is the topology involved rather than specific dynamical contents, and it can be applied to lower density even if the crystal structure may not be a good dynamical description there.

It has been established that at certain density, it is energetically more favorable that a skyrmion on FCC fractionize into two half-skrymions in CC or BCC~\cite{goldhaber,kugler}. What happens is that by squeezing the crystal size (increasing density), one induces a ``phase change"\footnote{I put this in parenthesis to indicate that it is not clear how to interpret the phenomenon in terms of the standard Ginzburg-Landau-Wilson paradigm for phase transitions. I will however loosely use this terminology for the changeover involved.} from skyrmion matter to half-skyrmion matter at a density  denoted  $n_{1/2}$~\cite{multifacet}. This phase change -- which is generic independently of the specific meson degrees of freedom involved apart from the pions -- engenders the change of the quark condensate defined on the average in the unit cell from $\Sigma\equiv \la\bar{q}q\ra|_{unit}\neq 0$ in the skyrmion phase to $\Sigma=0$  in the half-skyrmion phase. Locally the condensate is not equal to zero in the half-skyrmion phase, so there is a modulated scalar density distribution. In this phase, the pion decay constant does not vanish. Therefore, pions are propagating and so are other mesons and baryons. This means that although the $\Sigma$ is zero on the average, chiral symmetry is not actually restored, and the confinement persists. There must therefore be an order parameter for chiral symmetry that is of higher dimension field operators. This must be a medium-induced operator. One can also think of this phase as ``quarkyonic."

One may question whether this transition is real and not just an artifact of the crystal structure which may not be realistic at not so high density. I have no simple answer to this question since I do not know how to compute quantum corrections higher order in $1/N_c$. However I suggest that given that what is involved is a topology change involving different symmetries as in other areas of physics, the approach could well be reliable, unaffected qualitatively by higher order $1/N_c$ corrections. In fact, topology change is currently a deep issue in quantum physics~\cite{wilczek-topology} as well as a highly topical issue in condensed matter physics~\cite{wen}. What might be happening in the hadronic case I am dealing with is much less clear because we do not know what quantum ($1/N_c$) corrections will do. What I will do is to exploit the possibility that topology may be captured by a smooth change in boundary conditions in Hilbert space\footnote{An example that illustrates topology change in terms of change in boundary conditions is the chiral bag model~\cite{littlebag}. There the topological charge representing baryon charge can be continuously changed by  change in the bag boundary conditions.}  which in the present case, corresponds to  a change of parameters in the Lagrangian. The parameter change will then reflect the vacuum change in medium as density exceeds $n_{1/2}$.
\section{Parameter Changes from Nuclear Symmetry Energy}\label{symmetry-energy}
Since we do not know how to fully quantize skyrmion matter, the strategy is to extract density-dependent parameters of the effective Lagrangian capturing the topology change from the skyrmion crystal calculation and then do field theoretic many-body calculations with the Lagrangian so defined. This has been done by looking at the nuclear ``symmetry energy factor" $E_{sym}$~\cite{dongetal},
\be
E(n,\delta)=E_0(n)+E_{sym}(n)\delta^2+\cdots
\ee
where $E$ is the energy per baryon of the system, $E_0$ is the symmetric part of the energy with $\delta=0$ and $\delta=(N-Z)/(N+Z)$ with $N(Z)$ is the number of neutrons (protons). One can calculate the symmetry energy factor $E_{sym}$ by collective-quantizing the skyrmion crystal representing pure neutron matter~\cite{LPR}. One obtains
\be
E_{sym} (n)\approx \frac{1}{8\Omega_{cell}(n)}\label{SE}
\ee
where $\Omega_{cell}$ is the moment of inertial of the single cell of the crystal given by the integral over the cell with  the crystal configuration $U_0 (r)$ for the chiral field with the lowest energy for a given density.

\begin{figure}[h]
\centerline{
\includegraphics[width=0.35\textwidth,angle=-90]{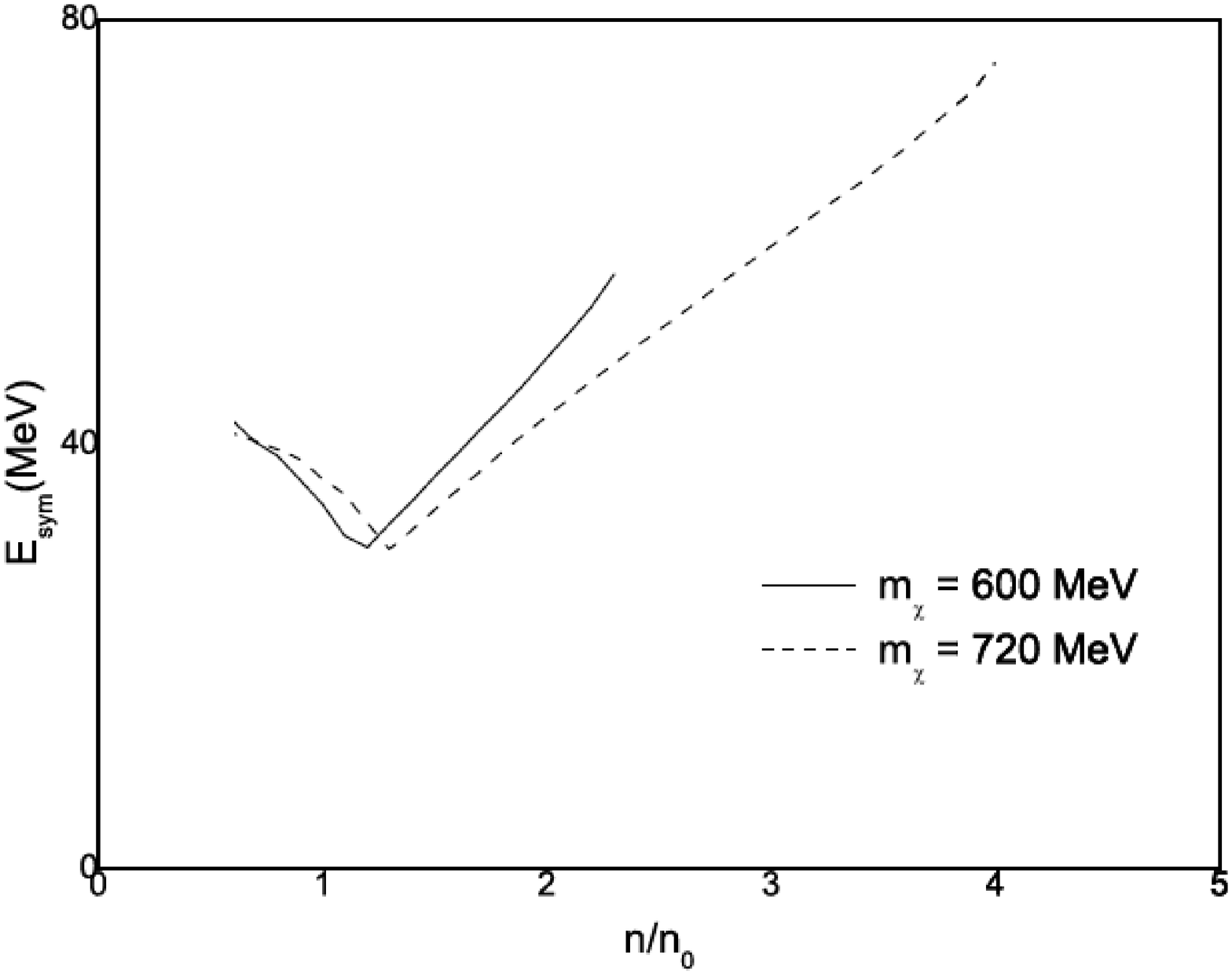}
}
\vskip -0.cm
\caption{Symmetry energy factor predicted by the skyrmion crystal at $n_{1/2}=1.3 n_0$. Note the cusp structure at the transition density.}
\label{skyrmeSE}
\end{figure}

The result is shown in Fig.~\ref{skyrmeSE} for two different masses for the dilaton appropriate for the system~\footnote{In relativistic mean field calculations of nuclei and nuclear matter, the scalar meson mass usually taken is $\sim 600$ MeV. This corresponds to the vacuum mass $\sim 750$ MeV which drops to $\sim 600$ MeV by scaling at nuclear matter density.}. The kinetic energy contribution -- suppressed for large $N_c$  -- is not included in (\ref{SE})\footnote{Because of the strong tensor forces, the kinetic energy contribution could in reality be strongly suppressed~\cite{tensors-BAL}.}. As I will show later, there are other correlation terms, higher order in $1/N_c$, that turn out to become significant in medium, so this cannot be compared  directly with nature. What is significant is the qualitative feature of the cusp at the transition density $n_{1/2}$.

To understand the cusp structure, it turns out to be most informative to look at the behavior of the tensor forces as density increases. This is because the symmetry energy is dominated by the tensor forces (for an updated account, see \cite{tensors-BAL}). Approximately, the symmetry energy factor $E_{sym}$ from the tensor forces in standard nuclear theory can be written as~\cite{BM}
\be
E_{sym}\sim \la V_{sym}\ra\approx \frac{12}{\bar{E}}\la V_T^2(r)\ra\label{BM}
\ee
where $\bar{E}\approx 200$ MeV is the average energy typical of the tensor force excitation and $V_T$ is the radial part of the net tensor force. This will be identified with the mean-field result of the Lagrangian with density-dependent parameters deduced from the topology change.

\begin{figure}[h]
\begin{center}
\includegraphics[height=5.5cm]{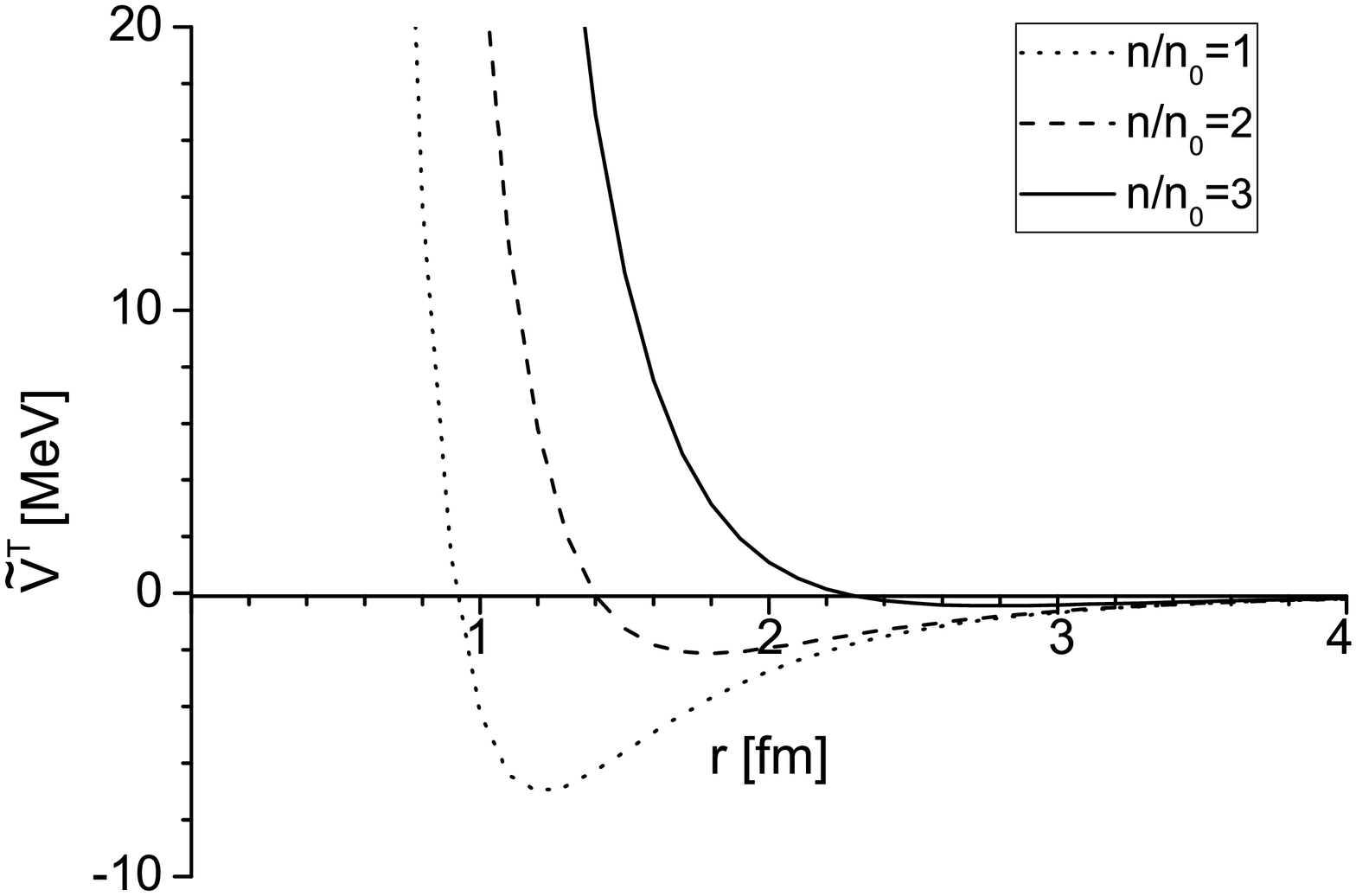}\, \includegraphics[height=5.5cm]{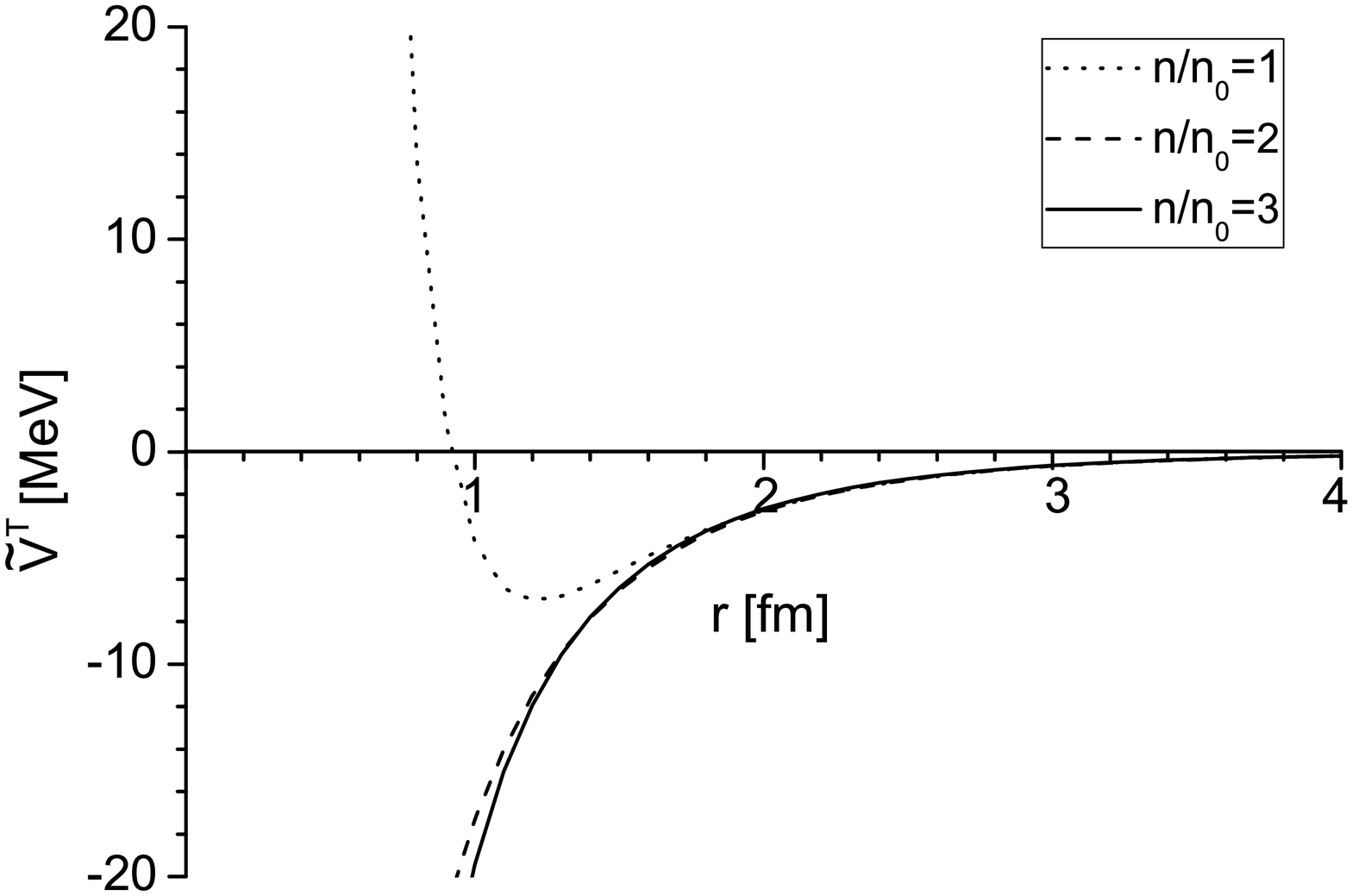}
\vskip -0.5cm
\caption{Sum of $\pi$ and $\rho$ tensor forces in units of MeV for densities $n/n_0$ =1, 2 and 3 with the ``old scaling" (upper panel) and with the ``new scaling," (lower panel) with $m_0\approx 0.8 m_N$. The topology change is put at $n_{1/2}=1.3n_0$.}\label{tensor}
\end{center}
\end{figure}

Described in the dHLS model adopted, the tensor forces between two nucleons (in principle obtained from skyrmions) are given by the exchange of a pion and a $\rho$ meson. They come in with opposite signs,  so that their effects tend to cancel. The property of their net effect depends sensitively on the scaling of the masses of the mesons and the nucleon involved as well as on the meson-nucleon coupling constants in medium. The main qualitative results from the skyrmion crystal combined with the VM/HLS and the dilaton-limit fixed point (DLFP for short) of dHLS discovered in \cite{DLFP} can be summarized by the modifications in the scaling of parameters in the Lagrangian from the ones proposed in 1991~\cite{BR91}. They  are, for $n\gsim n_{1/2}$,
\begin{itemize}
\item The effective mass of the nucleon as a soliton in medium goes like $m_N^*\sim f_\pi^*$ and the effective pion decay constant $f_\pi^*$ drops slowly in the half-skyrmion phase. Therefore the nucleon mass stays more or less un-scaling after $n_{1/2}$, consistent with a large $m_0$ in (\ref{Nmass}).
\item The effective coupling of the $\rho$ meson to nucleons $g_{\rho N}=g(1-g_V)$ where $g_V$ is the ``induced" vector coupling drops rapidly toward zero as density increases. This is because both $g\rightarrow 0$ due to the VM and $(1-g_V)\rightarrow 0$ due to the DLFP~\cite{DLFP}.
\end{itemize}
The net result is that the old scaling in the Lagrangian -- called BR scaling~\cite{BR91} -- is replaced by a new scaling at $n=n_{1/2}$~\cite{LPR,LR,dongetal}, that I will call  ``BLPR,"  standing for Brown, Lee, Park and Rho involved in various aspects of the scaling relation. One of the principal consequences of BLPR is that the $\rho$-tensor force is strongly suppressed for $n\geq n_{1/2}$, while the pion tensor remains more or less intact, thereby leaving only the $\pi$ tensor for $n\gsim 2n_0$. This is depicted in Fig.~\ref{tensor}, lower panel, for the case with $n_{1/2}\sim 1.3 n_0$ taken as an example\footnote{The exact location of $n_{1/2}$ is not determined from the crystal calculation performed so far. It will require an involved numerical work with the dHLS Lagrangian. With the simplified dSkyrme Lagrangian, it comes, for reasonable values of (scaling) parameters, to $1.3\lsim n_{1/2}/n_0\lsim 2.0$. This is the range supported by nuclear phenomenology~\cite{dongetal}.}.
We see from Fig.~\ref{tensor} that the net tensor strength at the intermediate and long ranges decreases due to the cancellation between the $\pi$ and $\rho$ tensors up to the density $n_{1/2}$, after which the $\rho$ tensor starts getting suppressed and the pion tensor quickly taking over with its full strength. It is easy to see that the cusp structure of the skyrmion result is reproduced qualitatively by the formula (\ref{BM}) with the BLPR scaling. This supports the identification of the scaling parameters with the topology change at the mean-field level. We will see that higher order nuclear correlations do smooth the cusp structure, but the main feature will remain in a more realistic treatment of EoS.

It should be noted that the change in the tensor forces, affecting importantly the symmetry energy of compact stars, could also have a precursor effect on the structure of asymmetric nuclei, such as for instance shell evolution, to be studied in RIB accelerators.
\section{Dilaton in nuclear matter}\label{dilaton}
In nuclear physics, it is essential that there be a scalar degree of freedom with a vacuum mass of $\sim 600-700$ MeV. It plays a key role in phenomenological as well as one-boson-exchange potentials and also in relativistic mean field approaches to nuclei and nuclear matter. Roughly, the scalar field provides the attraction to bind nucleons in nuclei with the stabilization provided by a vector repulsion. In nature there is no sharp and well-defined scalar in the vacuum with the mass needed. The meson $f_0(600)$ is listed in the particle data booklet, but its structure in terms of QCD is not understood and remains controversial despite intensive works on it and other scalars. In chiral perturbation theory, the scalar ``resonance" in $\pi-\pi$ interactions can be approximately reproduced by high-order loop calculations. Also the one-scalar-boson ($\sigma$) exchange in the OBEP NN potential can be approximately described by irreducible two-pion exchanges (involving $\Delta$ resonance intermediate states, form factors etc.). Both are perturbative in nature.

What we need for our purpose is, however, a local field  for the scalar excitation to be treated non-perturbatively, that is, in the mean-field approximation. One way to introduce such a scalar field in the context of scaling parameters in effective Lagrangians was first suggested in 1991~\cite{BR91}. What has been done in the work I am reporting here follows essentially that approach but in a much more refined and improved form. The basic idea is to link chiral symmetry to scale (or conformal) symmetry in such a way that {\it spontaneous breaking of scale symmetry (SBSS) triggers spontaneous breaking of chiral symmetry (SBCS)}. Scale symmetry cannot spontaneously break unless there is explicit breaking, and it is the trace anomaly that provides the natural source for the explicit breaking. It is precisely this observation that suggested the separation into ``soft glue" and ``hard glue" of the gluon condensate and the linking of the soft component to the dilaton condensate $\la\chi\ra$ which in turn is locked to the quark condensate$\la\bar{q}q\ra$.  (See \cite{LR-dilatons} for discussions on this matter and for previous references.)  One can associate this dilaton, i.e., soft glue, with the low-mass scalar needed in nuclei and nuclear matter. I note here an analogy to what is being done for the 125 GeV boson discovered at LHC, that is, to describe it as a Higgs-like dilaton, with scale symmetry broken spontaneously by a weak external conformal symmetry breaking. The pseudo-Goldstone nature of the boson is to account for the low mass of the discovered boson.

 Currently a highly topical issue, the QCD structure of the low-mass scalar is poorly known. It is most likely a complicated mixture of glue ball and $(q\bar{q})^n$ (with $n=1,2,\cdots$) configurations. For our purpose, we will not need to specify its detailed structure. We will simply take it to be the dilaton $\chi$ associated with spontaneous breaking of scale symmetry. Its condensate $\la\chi\ra$, locked to $\la\bar{q}q\ra$, will go to zero when chiral symmetry is restored (in the chiral limit).  The trace anomaly associated with the SBSS will then go to zero at the chiral transition. After chiral restoration there will still remain the explicit breaking due to the trace anomaly tied to the asymptotic freedom. This piece is chirally invariant and is presumably responsible for non-zero $m_0$.

In constructing the dHLS Lagrangian consistent with SBSS, the $\chi$ field enters as a ``conformal compensator field."  Except for the explicit chiral symmetry breaking term (i.e., pion mass term), the procedure is straightforward.  The condensate $\la\chi\ra$ -- signalling SBSS -- tracks the vacuum change as density is increased and makes the parameters of the Lagrangian dHLS change as the density of the system goes across $n_{1/2}$.

In the low density regime such as in nuclei and nuclear matter,  relativistic mean field models require that the $\chi$ field be chiral-singlet or dominated by a chiral-singlet component. However as one approaches the density at which chiral phase transition is to take place, the scalar should change over to the $\sigma$  -- which is the fourth component of the chiral four vector -- of the Gell-Mann-L\'evy sigma model or its generalization to the parity-doublet sigma model. How one can go from low density to high density in the framework of dHLS is discussed in terms of the dilaton limit fixed point (DLFP) in  \cite{DLFP}. The changeover presumably involves the scalar meson undergoing a sort of level-crossing at some density. It may also involve the role of $1/N_c$ corrections at varying densities. Both are very difficult questions to address at present.  Whether the picture we have adopted is viable or not remains to be checked by experiments.
\section{EoS of compact stars}
The change of the symmetry energy caused by the modified tensor forces at a density slightly above that of nuclear matter has an important consequence on EoS: the EoS which is soft below $n_{1/2}$,  gets hardened above $n_{1/2}$~\cite{LR,dongetal}. This is precisely the feature of EoS required for explaining massive compact stars with $M\sim 2 M_\odot$ being observed.  I would suggest that this feature could be checked in heavy-ion experiments that will probe baryon densities $n\gsim 2 n_0$.

Let me briefly describe, following \cite{dongetal},  how the approach given above fares in confronting compact stars. The idea here is to implement the BLPR in an effective field theory (EFT) for nuclear matter. Since the dHLS Lagrangian with the parameters fixed from the crystal calculation is a tree-level  Lagrangian, one needs to do quantum calculations with this Lagrangian to confront nature. In doing this, the ``double-decimation" strategy can be adopted~\cite{BR:DD,dongetal}.
\begin{enumerate}
\item The first decimation consists of obtaining via RG equation the $V_{lowk}$ in free space by decimating to the scale $\Lambda_{lowk}$ that describes nucleon-nucleon interactions up to lab momentum to $\sim 300$ MeV. It is best for our problem to do this in terms of the generalized HLS Lagrangian with the parameters with the intrinsic density dependence given by the BLPR.
\item
The second decimation is to do nuclear many-body calculation with this $V_{lowk}$ to the Fermi-momentum scale $\Lambda_{fermi}$. There are a variety of ways of doing this step. They all amount essentially to doing Landau Fermi-liquid fixed point theory and arrive at nuclear matter at the equilibrium density $n_0\sim 0.16$ fm$^{-3}$. This is an EFT well-justified up to density near $n_0$. This step fixes the scaling properties of the parameters in the Lagrangian up to near $n=n_0$. The same scaling is assumed up to $n_{1/2}$ provided it is not too high above $n_0$.
\item  The last step is to smoothly extrapolate with the formalism to high densities and calculate the EoS for compact stars. In doing this, one can adopt chiral perturbation strategy and include n-body forces -- if one wishes -- for $n>2$, suitably introducing scaling parameters for the n-body forces. In \cite{dongetal}, the topology change is incorporated in a smooth manner in terms of the changes in intrinsic scaling at $n_{1/2}$. It however ignores other degrees of freedom that might enter, such as strangeness that can be manifested in terms of kaon condensation (or equivalently hyperons) and strange quarks etc.
\end{enumerate}
\begin{figure}[here]
\scalebox{0.32}{\includegraphics[angle=-90]{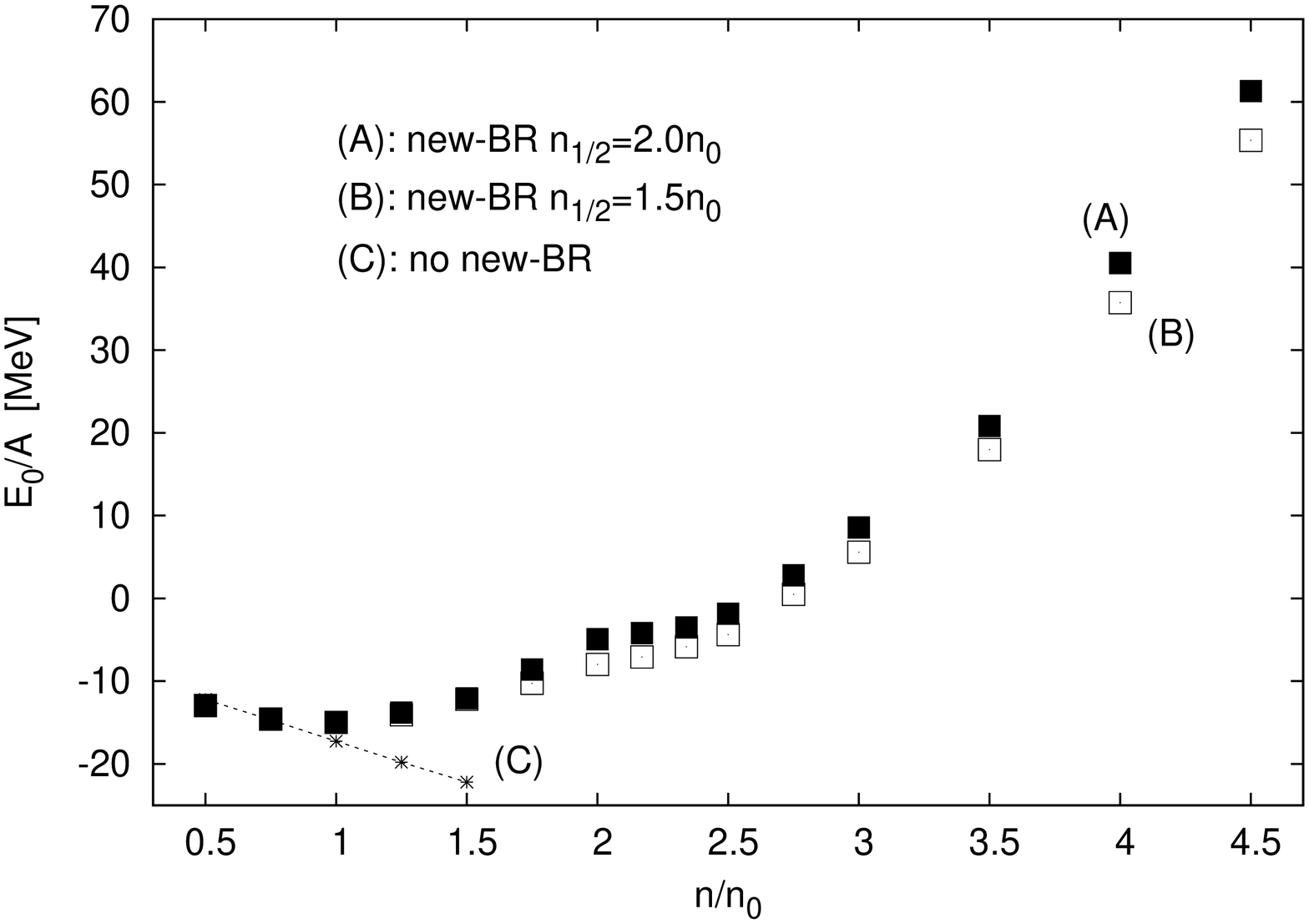}}
\scalebox{0.32}{\includegraphics[angle=-90]{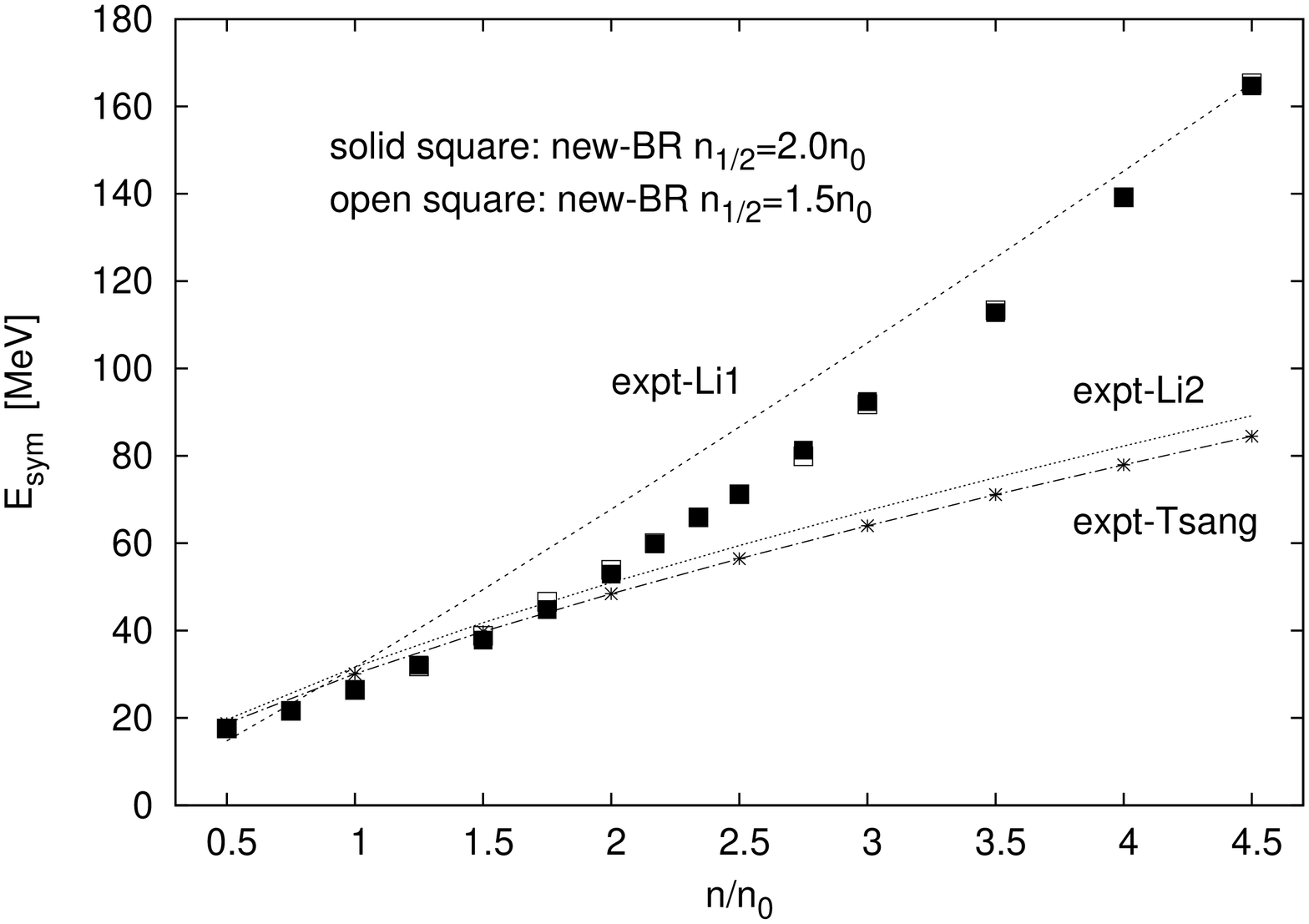}}
\caption{EoS for symmetric nuclear matter (upper panel) and symmetry energy (lower panel) with comparison with experimental fits,  for $n_{1/2}/n_0=1.5$ and $2.0$. }\label{figure3}
\end{figure}

The calculated results that come from the above procedures~\cite{dongetal} are given in Fig.~\ref{figure3}. The upper panel shows the EoS for symmetric nuclear matter and the lower panel the symmetry energy factor $E_{sym}$. One notes that without  BLPR, nuclear matter would saturate at too high a density with much too large a binding energy. For the symmetry energy, although the tree-order cusp is smoothed by higher-order nuclear correlations, it leaves a distinctive imprint in the change of its slope at $n_{1/2}$: it is soft below $n_{1/2}$ and becomes stiffer above $n_{1/2}$.
\begin{figure}[here]
\scalebox{0.32}{\includegraphics[angle=-90]{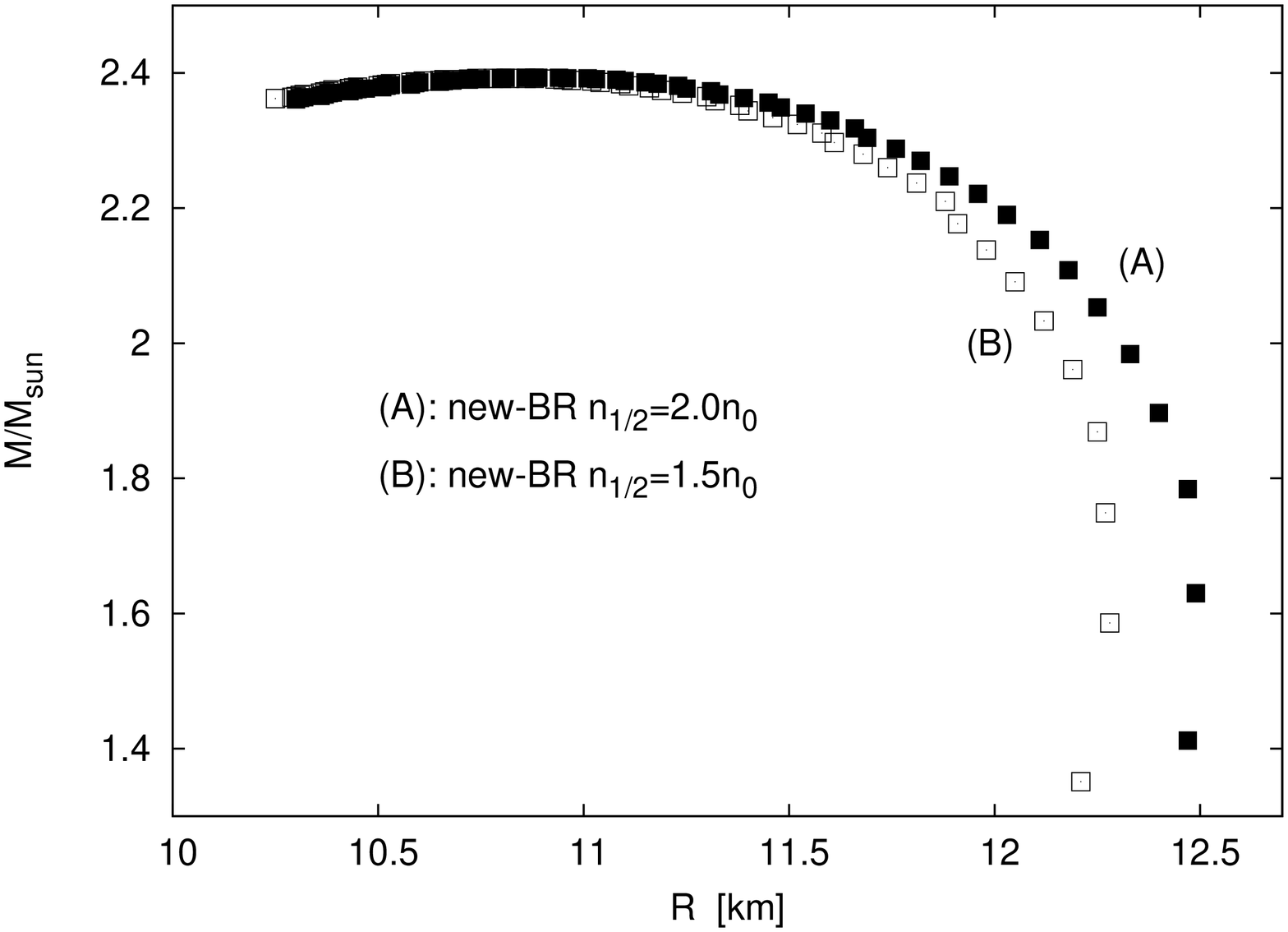}}
\scalebox{0.32}{\includegraphics[angle=-90]{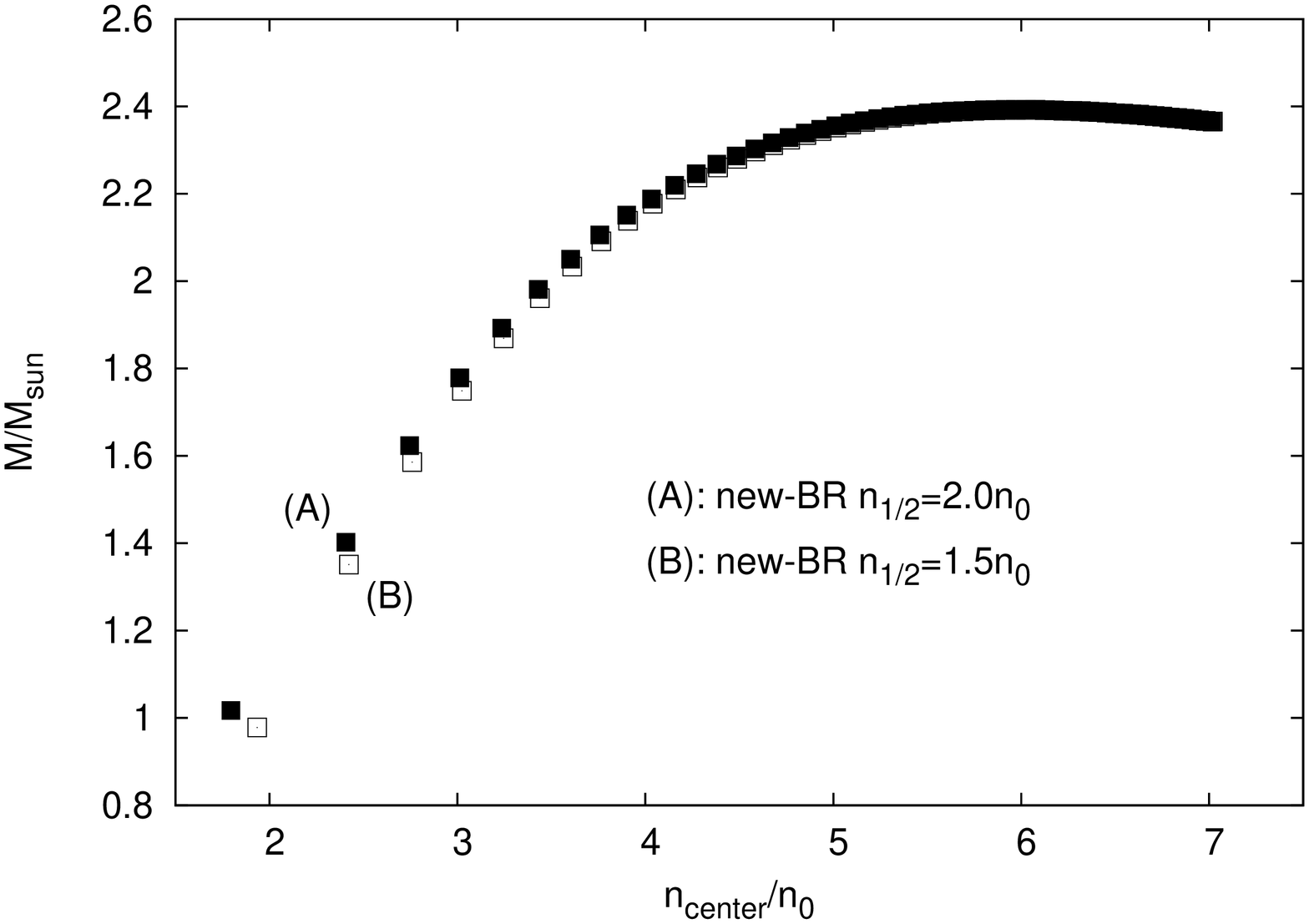}}
\caption{Mass-radius trajectories (upper panel) and central densities ($n_{center}$) (lower panel) of neutron stars calculated for $n_{1/2}$ = 2.0 (A) and 1.5$n_0$ (B).
The maximum neutron-star mass and its radius for these two cases are
respectively  (2.39 $M_{\odot}$, 10.90 km) and
(2.38 $M_{\odot}$,  10.89  km).}\label{stars}
\end{figure}

{\it Here is yet another case where it is dangerous to jump to a conclusion based on what is apparent. Similarly to the dilepton experiments and the anomalous orbital gyromagnetic ratio $\delta g_l$ of heavy nuclei where one could not naively associate what is observed with a  signal for chiral symmetry, here what is predicted from the topology change, i.e., the cusp, becomes nearly invisible in the background of many-body correlations.} We clearly need a cleverer idea to unearth the effect we would like to expose.

The equations of state so obtained predict the star properties depicted in Fig.~\ref{stars}. Thanks to the stiffening at $n_{1/2}$, one gets a massive star $M\sim 2.4 M_\odot$.  The maximum density reached for such a star is $\sim 5 n_0$.

\section {Strangeness degree of freedom}
If the central density of the massive compact star reaches $\sim 5 n_0$ as is found, then it is possible that kaons will condense or equivalently hyperons will appear.\footnote{It has been suggested in a unified approach with an effective Lagrangian that hyperons can appear {\it only when} kaons condense, which would mean that they represent the same physics. See \cite{hyperons-kaons}.} The appearance of the strangeness degrees of freedom will affect the EoS: It will soften it.  It was observed in \cite{PKR} that the onset of the half-skyrmion matter that leads to the stiffening of the EoS induced a propitious drop of the mass of the kaon propagating in dense medium. What this means in compact-star matter is that kaons will condense more rapidly with the kaon mass dropping faster. So there are two opposing phenomena taking place at $n_{1/2}$, one stiffening the EoS and the other softening. Since stiffening the EoS in the nucleon sector is known to push the kaon condensation to higher density, it is not at all clear what the net effect will be. This matter remains to be resolved. I should mention that condensed kaons do not necessarily imply that a massive star of $\sim 2 M_\odot$ cannot be formed. In fact, the state of condensed kaons can play the role of a doorway state to strange quark matter that can accommodate massive stars as suggested in \cite{LR,KLRastro}.
\section{Comments on the origin of hadron masses}
To conclude, I would like to make two comments.
\begin{enumerate}
\item One of the most important observations made in the work reported here is that in applying BLPR scaling to compact stars in \cite{dongetal}, one has no freedom to let the nucleon mass drop appreciably below $\sim 0.8 m_N$ for $n\gsim n_{1/2}$ with the other parameters held fixed. Lower nucleon mass would bring too strong a repulsion and make the symmetry energy -- and the EoS in general --  go haywire.  The question then is: Why can one not fiddle with other scaling parameters of the Lagrangian so as to compensate the effect of the dropping nucleon mass? For instance if the $\omega$-NN coupling in the dHLS Lagrangian is arbitrarily allowed to drop, then one may suitably soften the repulsion due to $\omega$ exchanges to compensate the repulsion from the decreased nucleon mass and keep the EoS within the range given by  heavy-ion experiments. However an approximate one-loop renormalization-group analysis made so far with the dHLS Lagrangian  indicates that the $\omega$-NN coupling does not scale~\cite{paengLR}, which means that at least at one-loop order, the coupling does not drop. This indicates that the $U(2)$ symmetry, fairly good in the (matter-free) vacuum for the $\rho$ and $\omega$, could be breaking down in medium,  given that the $\rho$ coupling {\em does flow} to the VM fixed point. Higher-loop RG analysis may be necessary to confirm this result. What would be the most exciting could be that the symmetry energy in compact stars resulting from the presence of half-skyrmion structure is pointing to a substantial $m_0$ which carries the crucial imprint of the origin of the proton mass.
\item Within the framework of dHLS, I have argued that while most of the nucleon mass need not follow the quark condensate (with a significant $m_0$), meson masses most likely do.  This conclusion would be invalidated if the constituent quark model which holds fairly well in the vacuum and which gets a strong theoretical support on the basis of large $N_c$ considerations~\cite{weinberg-pion}, held in dense matter. In this case, one could construct a parity-doublet model for the constituent quark instead of for the nucleon with a large chirally invariant mass $m_{0Q}$ with the mass formula for the constituent quark of the form $m_Q=m_{oQ}+\bar{m}_Q (\la\bar{q}q\ra)$. Then the mass ratio $m_M/m_B\simeq 2/3$ (where M and B stand, respectively, for light-quark mesons and baryons) which holds fairly well in matter-free space would hold equally well in the vicinity of the chiral restoration. This would be consistent with Glozman et al's observation that both baryons and mesons have rather large masses in the chirally restored phase~\cite{glozman}. However naive consideration paralleling the nucleon parity-doubler would not work since there will be problems with both meson and baryon spectra due to the parity doubling of the constituent quarks. A subtler approach will be needed to make that picture work, if at all\footnote{I would like to thank Leonid Glozman for helpful comments on this matter.}. Furthermore this scenario would seriously revamp both the ``old" and ``new" (BLPR) scalings.
\end{enumerate}
\subsection*{Acknowledgments}
This note is my personal account of work done in the WCUIII project at Hanyang University (partially supported by the Korean Ministry of Education, Science and Technology (R33-2008-000-10087-0)) with Hyun Kyu Lee, Kyungmin Kim and Won-Gi Paeng and with the participation of Masayasu Harada, Tom Kuo, Yong-Liang Ma, Yongseok Oh,  Byung-Yoon Park and Chihiro Sasaki. None of them should of course be held responsible for whatever errors I may be committing in my note. I am deeply grateful for discussions with all of them on the matters I presented, in particular with Hyun Kyu Lee on the role of topology and dilatons in hadron physics and Tom Kuo on doing realistic nuclear field theory calculations for compact stars.
\vskip 0.1cm



\begin{thebibliography}{50}

\bi{wilczek} F. Wilczek, ``Mass without mass," Physics Today 62N11 (1999) 11; Physics Today 53N11 (2000)13.

\bi{gell-mann-levy}  M.~Gell-Mann and M. L\'evy,
  ``The axial vector current in beta decay,''  Nuovo Cim.\  {\bf 16}, 705 (1960).  

\bi{detar} C.~E.~DeTar and T.~Kunihiro,
  ``Linear sigma model with parity doubling,''  Phys.\ Rev.\ D {\bf 39}, 2805 (1989).  

\bi{HY:PR}    M.~Harada and K.~Yamawaki, Phys.\ Rept.\  {\bf 381}, 1 (2003).

\bi{weinberg} S.~Weinberg,
  ``Mended symmetries,''  Phys.\ Rev.\ Lett.\  {\bf 65}, 1177 (1990);  
 ``Unbreaking symmetries,''  Conf.\ Proc.\ C {\bf 930308}, 3 (1993).  

  \bi{Lang-unbreaking} C.~B.~Lang and M.~Schr\"ock,
  ``Unbreaking chiral symmetry,''  Phys.\ Rev.\ D {\bf 84}, 087704 (2011)  [arXiv:1107.5195 [hep-lat]].  

\bi{glozman} L.~Y.~Glozman,
  ``Confinement, chiral symmetry breaking and the mass generation of hadrons,''  arXiv:1211.7267 [hep-ph];  
L.~Y.~.Glozman, C.~B.~Lang and M.~Schr\"ock,
  ``Symmetries of hadrons after unbreaking the chiral symmetry,''  Phys.\ Rev.\ D {\bf 86}, 014507 (2012)  [arXiv:1205.4887 [hep-lat]]. 





\bi{blind}  M.~Rho,
  ``Dileptons get nearly 'blind' to mass-scaling effects In hot and/or dense matter,''  arXiv:0912.3116 [nucl-th].  

\bibitem{BR91} G.~E.~Brown and M.~Rho,
  ``Scaling effective Lagrangians in a dense medium,''
  Phys.\ Rev.\ Lett.\  {\bf 66}, 2720 (1991).
\bi{KR}  K.~Kubodera and M.~Rho,
  ``Axial charge transitions in heavy nuclei and in-medium effective chiral Lagrangians,''  Phys.\ Rev.\ Lett.\  {\bf 67}, 3479 (1991).  


\bi{friman-rho}  B.~Friman and M.~Rho,
  ``From chiral Lagrangians to Landau Fermi liquid theory of nuclear matter,''  Nucl.\ Phys.\ A {\bf 606}, 303 (1996)  [nucl-th/9602025].  

\bi{maetal} Y.~-L.~Ma et al,
  ``Hidden local symmetry and infinite tower of vector mesons for baryons,''  Phys.\ Rev.\ D {\bf 86}, 074025 (2012)  [arXiv:1206.5460 [hep-ph]];  
Y.~-L.~Ma, G.~-S.~Yang, Y.~Oh and M.~Harada,
  ``Skyrmions with vector mesons in the hidden local symmetry approach,''  arXiv:1209.3554 [hep-ph];  
Y.-L. Ma, M. Harada, H.K. Lee, B.-Y. Park, Y. Oh and M. Rho, work in progress.





 \bi{skyrme}  T.H.R. Skyrme,
``A unified field theory of mesons and baryons,"
{Nucl. Phys.} {\bf 31}, 556 (1962).


\bibitem{multifacet} B.-Y. Park and V. Vento, ``Skyrmion approach to finite density and temperature" in  {\it The Multifaceted Skyrmion}\ (World Scientific, Singapore, 2010) ed. G.E. Brown and M. Rho.

\bi{klebanov} I.~R.~Klebanov,
  ``Nuclear matter In the Skyrme model,''
  Nucl.\ Phys.\ B {\bf 262}, 133 (1985).  

\bi{goldhaber} A.~S.~Goldhaber and N.~S.~Manton,
  ``Maximal symmetry of the Skyrme crystal,''  Phys.\ Lett.\ B {\bf 198}, 231 (1987).  

\bi{kugler}  M.~Kugler and S.~Shtrikman,
  ``Skyrmion crystals and their symmetries,''  Phys.\ Rev.\ D {\bf 40}, 3421 (1989).  


\bi{wilczek-topology} A.~D.~Shapere, F.~Wilczek and Z.~Xiong,
  ``Models of topology change,''  arXiv:1210.3545 [hep-th].  

\bi{wen}  L.~-Y.~Hung and X.~-G.~Wen,
  ``Quantized topological terms in weakly coupled gauge theories and their connection to symmetry protected topological phases,''  arXiv:1211.2767 [cond-mat.str-el].  

\bi{littlebag} G.~E.~Brown and M.~Rho,
  ``The little bag,''  Phys.\ Lett.\ B {\bf 82}, 177 (1979);  
M.~Rho, A.~S.~Goldhaber and G.~E.~Brown,
  ``Topological soliton bag model for baryons,''  Phys.\ Rev.\ Lett.\  {\bf 51}, 747 (1983).  

\bi{dongetal}  H.~Dong,
T.~T.~S.~Kuo, H.~K.~Lee, R.~Machleidt and M.~Rho,
``Half-skyrmions and the equation of state for compact-star matter,''
  arXiv:1207.0429 [nucl-th].  

\bi{LPR}   H.~K.~Lee, B.~Y.~Park and M.~Rho,
  ``Half-skyrmions, tensor forces and symmetry energy in cold dense matter,''
  Phys.\ Rev.\  C {\bf 83}, 025206 (2011).

\bi{tensors-BAL}  B.~-A.~Li, L.~-W.~Chen, F.~J.~Fattoyev, W.~G.~Newton and C.~Xu,
  ``Probing nuclear symmetry energy and its imprints on properties of nuclei, nuclear reactions, neutron stars and gravitational waves,''  arXiv:1212.1178 [nucl-th].  


\bibitem{BM}  G.~E.~Brown and R.~Machleidt,
  ``Strength of the $\rho$ meson coupling to nucleons,''
  Phys.\ Rev.\ C\ {\bf 50}, 1731  (1994).

\bi{DLFP}   W.~-G.~Paeng, H.~K.~Lee, M.~Rho and C.~Sasaki,
  ``Dilaton-limit fixed point in hidden local symmetric parity doublet model,''
  Phys.\ Rev.\ D {\bf 85}, 054022 (2012);
    C.~Sasaki, H.~K.~Lee, W.~-G.~Paeng and M.~Rho,
  ``Conformal anomaly and the vector coupling in dense matter,''
  Phys.\ Rev.\ D {\bf 84}, 034011 (2011).

  \bibitem{LR} H.~K.~Lee and M.~Rho,
  ``Flavor symmetry and topology change in nuclear symmetry energy for compact stars,''
  arXiv:1201.6486 [nucl-th].  



\bi{LR-dilatons} H.~K.~Lee and M.~Rho,
  ``Dilatons in hidden local symmetry for hadrons in dense matter,''  Nucl.\ Phys.\ A {\bf 829}, 76 (2009)  [arXiv:0902.3361 [hep-ph]].  

\bi{BR:DD}  G.~E.~Brown and M.~Rho,
  ``Double decimation and sliding vacua in the nuclear many body system,''
  Phys.\ Rept.\  {\bf 396}, 1 (2004)
  [nucl-th/0305089].

  \bi{hyperons-kaons} H.K. Lee and M. Rho, ``Hyperons and condensed kaons in compact stars," to appear.

\bi{PKR} B.~-Y.~Park, J.~-I.~Kim and M.~Rho,
  ``Kaons in dense half-skyrmion matter,''  Phys.\ Rev.\ C {\bf 81}, 035203 (2010)  [arXiv:0912.3213 [hep-ph]].  
  
\bi{KLRastro} K.~Kim, H.~K.~Lee and M.~Rho,
  ``Dense stellar matter with strange quark matter driven by kaon condensation,''  Phys.\ Rev.\ C {\bf 84}, 035810 (2011)  [arXiv:1102.5167 [astro-ph.HE]].  






\bi{paengLR} W.-G. Paeng, H.K. Lee and M. Rho, work in progress.

\bi{weinberg-pion} S.~Weinberg,
  ``Pions in large-$N$ quantum chromodynamics,''  Phys.\ Rev.\ Lett.\  {\bf 105}, 261601 (2010)  [arXiv:1009.1537 [hep-ph]].  

\end{thebibliography}
\end{document}